\documentclass[twocolumn,aps,prl,superscriptaddress,nofootinbib]{revtex4-2}
\usepackage{graphicx}
\usepackage{color}
\graphicspath{{figures/}{fig/}}
\usepackage{amsmath}
\usepackage{amssymb}
\usepackage{bm}
\usepackage{slashed}
\usepackage{epsfig}
\usepackage{amsfonts}
\usepackage{epstopdf}
\usepackage{hyperref}
\newcommand{\sect}[1]{{\it \textbf{#1.} --- }}

\begin{document}
\title{Electroweak Corrections to Double Higgs Production at the LHC}

\author{Huan-Yu Bi}
\email{bihy@pku.edu.cn}
\affiliation{School of Physics, Peking University, Beijing 100871, China}
\affiliation{Center for High Energy Physics, Peking University, Beijing 100871, China}

\author{Li-Hong Huang}
\email{lhhuang@pku.edu.cn}
\affiliation{School of Physics, Peking University, Beijing 100871, China}

\author{Rui-Jun Huang}
\email{huangrj2215@pku.edu.cn}
\affiliation{School of Physics, Peking University, Beijing 100871, China}

\author{Yan-Qing Ma}
\email{yqma@pku.edu.cn}
\affiliation{School of Physics, Peking University, Beijing 100871, China}
\affiliation{Center for High Energy Physics, Peking University, Beijing 100871, China}

\author{Huai-Min Yu}
\email{yuhm@stu.pku.edu.cn}
\affiliation{School of Physics, Peking University, Beijing 100871, China}

\date{\today}

\begin{abstract}
We present the results for the complete next-to-leading order electroweak corrections to $pp \to HH$ at the Large Hadron Collider, focusing on the dominant gluon-gluon fusion process. While the corrections at the total cross-section level are approximately $-4\%$, those near the energy of $HH$ production threshold exceed $+15\%$, and corrections at the high-energy region are around $-10\%$, leading to a shape distortion for the differential distributions. Our findings substantially diminish the theoretical uncertainties associated with this pivotal process,  providing valuable input for understanding the shape of the Higgs boson potential upon comparison with experimental measurements.
\end{abstract}

\maketitle
%\allowdisplaybreaks

%%%%%%%%%%%%%%%%%%%%%%%%%%%%%%%%%%%%%%%%%%%%%%%%%

\sect{Introduction}
After the groundbreaking discovery of the Higgs boson \cite{ATLAS:2012yve,CMS:2012qbp}, investigating the properties of the Higgs particle has emerged as a paramount avenue for unraveling the intricacies of electroweak (EW) symmetry breaking and the standard model (SM). One of the main focuses nowadays at the Large Hadron Collider (LHC) is the meticulous examination of Higgs self-interactions, as this endeavor is pivotal in elucidating the elusive shape of the Higgs boson potential, which remains unknown.

The exploration of Higgs boson pair production, a process intricately tied to the Higgs trilinear coupling $\lambda_{HHH}$, presents a unique opportunity for advancing our comprehension in this domain. With the continuous accumulation of data at the LHC and the potential enhancements in experimental techniques, there is a heightened expectation that $\lambda_{HHH}$ will be constrained to a reasonable range \cite{ATLAS:2022jtk,CMS:2022dwd,Cepeda:2019klc}.
The necessity for the modification of the Higgs boson potential arises if the measured value deviates from the SM value.

Theoretically, gluon-gluon fusion~\cite{DiMicco:2019ngk,Baglio:2012np,Frederix:2014hta} stands as the predominant production mode for Higgs pairs at the LHC, with rates for alternative production modes being at least an order of magnitude smaller. The leading-order (LO) amplitude for $gg\to HH$ is loop induced~\cite{Eboli:1987dy,Glover:1987nx,Plehn:1996wb}, owing to the absence of interactions between the Higgs boson and gluons in the SM. This lack of interaction makes fixed-order perturbative calculations highly challenging. The comprehensive next-to-leading order (NLO) QCD calculation was accomplished only a few years ago~\cite{Borowka:2016ehy,Borowka:2016ypz,Baglio:2018lrj,Davies:2019dfy}, and subsequent to this, the resummation of soft gluons and parton shower effects was incorporated~\cite{Ferrera:2016prr,Heinrich:2017kxx,Jones:2017giv,Heinrich:2019bkc}.
In the heavy top-quark mass limit approximation, the next-to-next-to-next-to-leading order ($\rm N^3LO$) QCD corrections have been obtained \cite{Chen:2019lzz}, coupled with soft-gluon threshold resummation \cite{AH:2022elh}. This achievement yields a scale uncertainty of approximately $1.7\%$, with additional uncertainties stemming from our imprecise knowledge of parton distribution functions (PDFs) and $\alpha_s$, accounting for about $3\%$ of the total uncertainty \cite{Jones:2023uzh}.

NLO EW corrections are notably significant in the high-energy region, primarily attributed to the Sudakov effect \cite{Sudakov:1954sw,Denner:2000jv,Ciafaloni:1998xg}, which has been demonstrated by many processes. For example, $-10\%$ to $-30\%$ corrections are observed in the high invariant mass or high transverse momentum region for vector-boson pair production at the LHC \cite{Bierweiler:2013dja}. Understanding the impact of NLO EW corrections on Higgs pair production, which represents the most substantial source of uncertainties on the theoretical front, is imperative. Additionally, the Higgs quartic coupling, another crucial quantity defining the Higgs boson potential, only emerges at the NLO EW level. Consequently, computing NLO EW corrections for $gg \to HH$ has been a focal point in the 2015, 2017, 2019, and 2021 Les Houches precision wish lists \cite{Andersen:2016qtm,Proceedings:2018jsb,Amoroso:2020lgh,Huss:2022ful}.

Calculating the NLO EW corrections to  $gg \to HH$ is inherently more intricate than computing NLO QCD corrections, primarily due to the challenging two-loop Feynman integrals involving a multitude of mass scales. Several attempts to address the EW corrections have been made in the literature \cite{Borowka:2018pxx,Muhlleitner:2022ijf,Davies:2022ram,Davies:2023npk,talk,talk2}, though they are either partial results or applicable only in specific regions.
For instance, NLO EW corrections are computed using heavy top-quark mass expansion in Ref. \cite{Davies:2023npk}, revealing suboptimal convergent behavior in the physical region and corrections as substantial as $65\%$ when the partonic center-of-mass energy is around $260$ GeV. 

These findings underscore the critical need for a comprehensive computation of the NLO EW corrections to $gg \to HH$ to ensure a reliable prediction. 
This has been addressed in this Letter, encompassing all Feynman diagrams at the two-loop level and incorporating all mass effects.

\begin{figure}[htbp]
%\begin{center}
    \includegraphics[width=0.4\textwidth, trim = 25 12cm 25 0cm, clip]{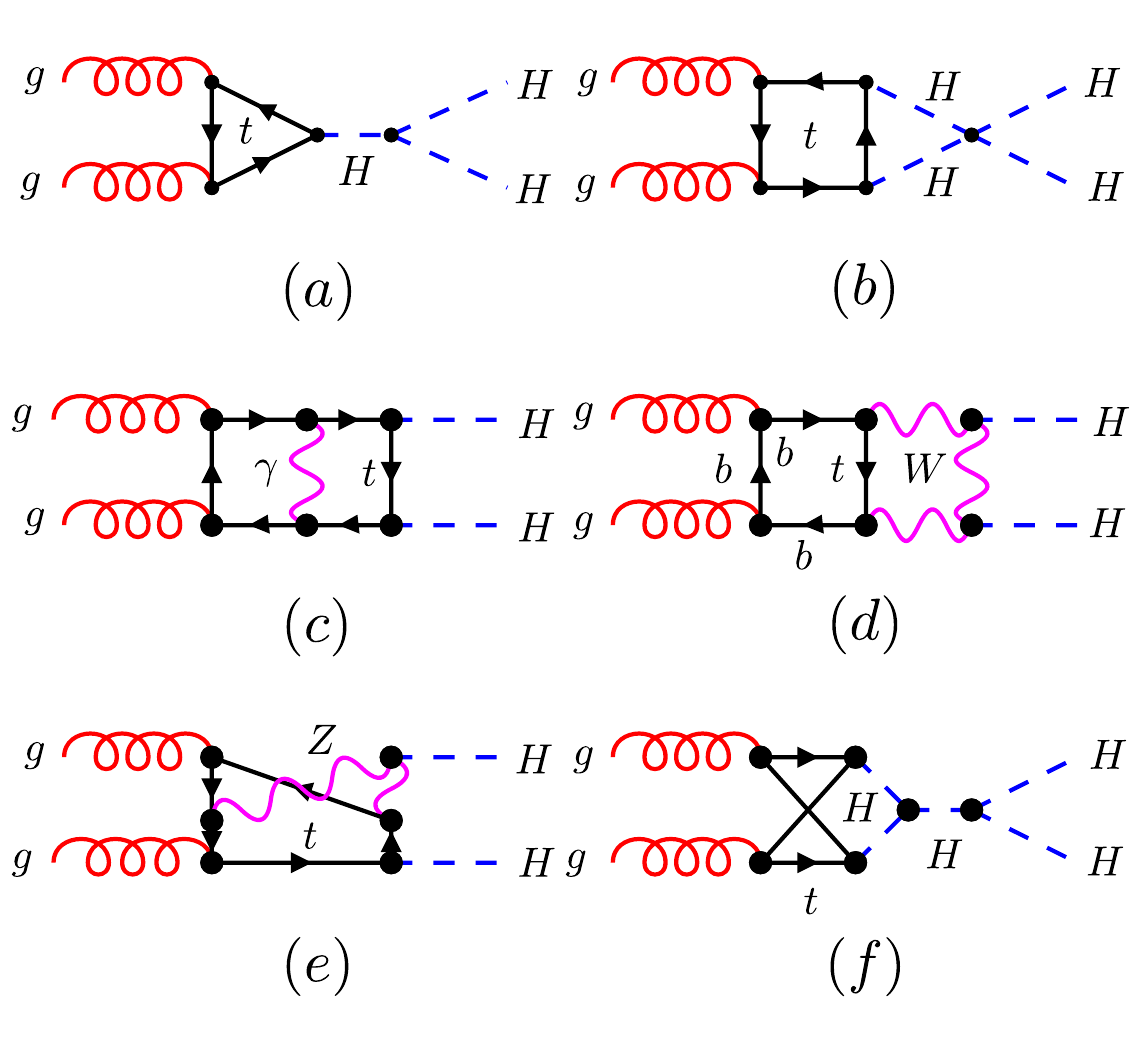}
\caption{\label{fig:feynmandiagrams}
Representative Feynman diagrams for $gg\to HH$ at LO (a) and NLO EW corrections (b)-(f). }
%\end{center}
\end{figure}

\sect{Amplitudes}
NLO EW corrections typically encompass both real-emission and virtual contributions. However, in the context of the current problem, the process involving one additional photon emission, $gg\to HH\gamma$, is prohibited by the Furry theorem. Additionally, heavy EW particle emissions can be experimentally distinguished. Consequently, NLO EW corrections for $gg\to HH$ exclusively involve virtual contributions, specifically at the two-loop level. The generation of Feynman diagrams and amplitudes is facilitated by {\tt FeynArt} \cite{Hahn:2000kx}, with some representative Feynman diagrams illustrated in Fig.~\ref{fig:feynmandiagrams}.

The amplitude for $g(p_1)g(p_2)\to H(p_3)H(p_4)$, ${\cal M}=\varepsilon_{1\mu}\varepsilon_{2\nu}{\cal M}^{\mu\nu}$ with $\varepsilon_{i\mu}$ denoting gluon polarization vector, satisfies the current conservation relations $p_{1\mu}{\cal M}^{\mu\nu}=p_{2\nu} {\cal M}^{\mu\nu}=0$. We have the following decomposition with each term satisfying the same relations:
\begin{align}
{\cal M}^{\mu\nu}=F_1T_1 ^{\mu\nu}+F_2T_2 ^{\mu\nu}+\Delta_0^{\mu\nu}+\Delta_5^{\mu\nu},
\end{align}
where $\Delta_5^{\mu\nu}$ is linearly dependent on the Levi-Civita tensor, $\Delta_0^{\mu\nu}$ depends on either $p_1^\mu$ or $p_2^\nu$,  $T_i ^{\mu\nu}$ contain other tensor structures which only have two degrees of freedom, and $F_i$ are gauge invariant form factors. As $\Delta_5^{\mu\nu}$ appears only from the two-loop order onward, its contribution vanishes for NLO EW calculations when multiplied by the LO amplitude. The contribution from $\Delta_0^{\mu\nu}$ vanishes due to current conservation relations. The remained tensors  $T_1 ^{\mu\nu}$ and $T_2 ^{\mu\nu}$ can be chosen as \cite{Plehn:1996wb}
\begin{align}
&T_1 ^{\mu\nu}= g^{\mu\nu}-\frac{p_1^{\nu}\,p_2^{\mu}}{p_1\cdot  p_2} \;,\label{eq:Ttensors} \\
&T_2 ^{\mu\nu}= g^{\mu\nu}+\frac{1}{p_{T}^{2}\,(p_1\cdot p_2)}\big[ 2\,(p_1\cdot p_2) \,p_3^{\nu}\,p_3^{\mu}\nonumber  \\
&~~ - 2\,(p_1\cdot p_3) \,p_3^{\nu}\,p_2^{\mu} -2\,(p_2\cdot p_3) p_3^{\mu}\,p_1^{\nu}+ m_H^2 \,p_1^{\nu}\,p_2^{\mu} \big] \;, 
\end{align}  
where $p_{T}=\sqrt{(\hat{t}\hat{u} - m_H^4)/\hat{s}}$ represents the transverse momentum of one of the Higgs
%\footnote{Since the two Higgs bosons are identical, we do not distinguish between the hardest and the second hardest Higgs.} 
with
\begin{eqnarray}
\hat{s}=(p_1+p_2)^2,\quad\hat{t}=(p_1-p_3)^2,\quad\hat{u}=(p_2-p_3)^2.\quad\label{st}
\end{eqnarray}   
Since $T_1\cdot T_2=D-4$ and $T_1\cdot T_1= T_2\cdot T_2=D-2$, where $D$ represents the generic spacetime dimension used to regularize potential divergences, we define the projectors
\begin{eqnarray}
P_1^{\mu\nu} &=&+\frac{1}{4}\,\frac{D-2}{D-3} \,T_1^{\mu\nu}
-\frac{1}{4}\,\frac{D-4}{D-3} \,T_2^{\mu\nu}\;,\label{eq:proj1}\\
P_2^{\mu\nu}&=& -\frac{1}{4}\,\frac{D-4}{D-3} \,T_1^{\mu\nu}
+\frac{1}{4}\,\frac{D-2}{D-3} \,T_2^{\mu\nu}\;,\label{eq:proj2}
\end{eqnarray}   
such that 
\begin{eqnarray}
F_1=P_1^{\mu\nu} {\cal M}_{\mu\nu},\quad F_2=P_2^{\mu\nu} {\cal M}_{\mu\nu}.
\end{eqnarray}

Because $\Delta_5^{\mu\nu}$ contributes nothing to NLO EW corrections, an even number of $\gamma_5$ terms is required in each contribution to the form factors. Additionally, we observe that terms containing $\gamma_5$ make no contribution in diagrams involving two fermion loops. Consequently, the even number of $\gamma_5$ terms can only arise in a single fermion loop, making the choice of the simplest naive $\gamma_5$ scheme, adopting the anticommutativity relation $\left\{ \gamma_5,\gamma_{\mu}\right\}=0 $ naively, unambiguous in this context.

Utilizing the {\tt {CalcLoop}} package \cite{calcloop}, the form factors are expressed as linear combinations of scalar Feynman integrals, which are categorized into 3 (116) integral families based on the type of propagators at the one-loop (two-loop) level. The loop integrals in each family are then further reduced to a more manageable basis, referred to as master integrals, through the utilization of {\tt{Blade}} \cite{Guan:2024byi}. {\tt{Blade}} harnesses the power of {\tt{FiniteFlow}} \cite{Peraro:2019svx} to solve integration-by-parts relations \cite{Chetyrkin:1981qh}, and it is equipped with the block-triangular form \cite{Liu:2018dmc, Guan:2019bcx} to further enhance computational efficiency (for other reduction packages on the market, see Refs.~\cite{Anastasiou:2004vj, Smirnov:2008iw,Studerus:2009ye,vonManteuffel:2012np,Lee:2012cn, Smirnov:2013dia,Lee:2013mka,Smirnov:2014hma, Maierhofer:2017gsa, Maierhofer:2018gpa,Smirnov:2019qkx,Klappert:2020nbg,Wu:2023upw}). 
As the final physical result is finite and insensitive to a small dimensional regulator $\epsilon$ with $D=4-2\epsilon$, we set $\epsilon=1/1000$ from the beginning to the end of the calculation, aligning with the approach proposed in Refs.~\cite{Liu:2022chg,Liu:2022mfb,Chen:2022vzo}. This choice avoids dealing with Laurent expansion with respect to $\epsilon$ in intermediate steps, thereby significantly reducing computational resources. 
Through computations utilizing an alternative point $\epsilon=-1/1000$, we observe that the outcomes deviate by a maximum of 0.4\% from those obtained with $\epsilon=1/1000$. This serves as a confirmation of the effective cancellation of divergences. By combining the finite components of both sets of results, we can further mitigate the error to $\mathcal{O}(\epsilon^2)$.

Master integrals are evaluated by numerically solving differential equations with respect to kinematical variables $\hat{s}$ and $\hat{t}$  \cite{Kotikov:1990kg,Remiddi:1997ny,Caffo:2008aw,Czakon:2008zk,Henn:2013pwa,Lee:2014ioa,Mandal:2018cdj,Moriello:2019yhu,Hidding:2020ytt,Armadillo:2022ugh} with boundary conditions provided by the  {\tt AMFlow} method \cite{Liu:2022chg,Liu:2022mfb,Liu:2021wks,Liu:2022tji,Liu:2020kpc,Liu:2017jxz}, as described in Ref.~\cite{Liu:2023jkr}. Singularities in the master integrals occur when intermediate particles go on shell, corresponding to $\sqrt{\hat{s}}=2m_H,m_W+m_t,2m_t,2m_t+m_Z$ or $2m_t+m_H$ in the physical region. Analytical continuation is executed by introducing an infinitesimal positive imaginary part to $\hat{s}$ across these singularities. Armed with this information, we can compute master integrals at any phase space point starting from the nearest computed point by solving the differential equations with the assistance of the differential solver in {\tt AMFlow} \cite{Liu:2022chg}. To validate our approach, we compared the results of all master integrals obtained in this manner against direct evaluations using the {\tt AMFlow} method at several phase space points. We observed at least 20 digits of agreement, ensuring the robustness and accuracy of our calculations.

For the renormalization of masses and fields, we employ the on-shell scheme, while the renormalization of the electromagnetic coupling $\alpha$ is conducted in the $G_{\mu}$ scheme \cite{Denner:2019vbn}. Following the renormalization process, we obtain finite results for $F_1$ and $F_2$, and we have explicitly verified the cancellation of divergences at several phase space points. 

Among the partial results reported in the literature, those presented in Refs. \cite{Davies:2023npk,talk} can be readily compared with our own findings. In the case of Ref. \cite{Davies:2023npk}, by choosing a nonphysical top-quark mass of 244.22 GeV to ensure the convergence of the heavy top-quark mass expansion, our results exhibit agreement to within approximately 0.1\%. However, when considering the physical top-quark mass, our comprehensive calculations reveal corrections of about 34\% at a chosen phase space point, as compared to the 65\% corrections reported in Ref. \cite{Davies:2023npk} for the partial result. Regarding the partial results obtained in Ref. \cite{talk}, our corresponding findings agree to within approximately 0.5\%. This minor discrepancy is primarily attributed to differences in the input parameters used in each study.

\sect{Total cross sections}
The cross sections of gluon-gluon fusion channel for $pp\to HH$ at LO or NLO can be expressed as
\begin{align}\label{eq:Xsection}
\sigma^{\text{LO(NLO)}} &= \frac{1}{512\pi}\int_0^1 \mathrm{d}x_{1}\int_0^1 \mathrm{d}x_{2} \int^{\hat{t}_{+}}_{\hat{t}_{-}}\mathrm{d}\hat{t} \nonumber \\
& \times \frac{1}{\hat{s}^{2}}f_{g/p}(x_{1},\mu)f_{g/p}(x_{2},\mu) \hat{\sigma}^{\text{LO(NLO)}},
\end{align}
where $\hat{t}^{\pm}=m^{2}_{H}-\frac{\hat{s}}{2}(1\mp\sqrt{1-{4m^{2}_{H}}/{\hat{s}}})$, $f_{g/p}(x,\mu)$ denotes the gluon distribution function inside of the proton, $\mu$ represents the factorization scale, and $\hat{\sigma}^{\text{LO(NLO)}}$ are given by
\begin{align}
\hat{\sigma}^{\text{LO}}&=\sum_{i=1}^2 |F_{i}^{(0)}|^2 \;,\\
\hat{\sigma}^{\text{NLO}}&=\sum_{i=1}^2 |F_{i}^{(0)}|^2+ F_{i}^{(0)}F_{i}^{(1)*}+F_{i}^{(0)*}F_{i}^{(1)}\; ,
\end{align}
where $F_i^{(0)}$ and $F_i^{(1)}$ correspond to the lowest order and the next-order terms in the $\alpha$ expansion of the form factors.

We use the following SM parameters, 
\begin{eqnarray}
\frac{m_H^2}{m_t^2}=\frac{12}{23},\quad\frac{m_Z^2}{m_t^2}=\frac{23}{83},\quad\frac{m_W^2}{m_t^2}=\frac{14}{65},
\end{eqnarray}
and the top-quark mass is chosen as $m_t=172.69$  GeV \cite{ParticleDataGroup:2022pth}, keeping all other masses and widths to zero.  This approximation may result in a slight deviation of a few percent in the determination of the NLO EW corrections, which is, however, insignificant for our intended purposes. $\alpha$ is calculated from the Fermi constant $G_{\mu}$ via
\begin{eqnarray}
\alpha=\frac{\sqrt{2}}{\pi}G_{\mu}m_W^2 \left(1-\frac{m_W^2}{m_Z^2}\right),
 \end{eqnarray}
which results in $\alpha=1/133.12=7.512\times 10^{-3}$ with input $G_{\mu}=1.166378\times 10^{-5}~{\rm GeV}^{-2}$.
The Cabibbo-Kobayashi-Maskawa mixing matrix is set to diagonal. We employ NNPDF3.1 \cite{NNPDF:2017mvq} as our default PDF set, with $\rm NNPDF31\underline{~}nlo\underline{~}as\underline{~}0118$ for both LO and NLO calculations. The running of strong coupling $\alpha_s$ with two-loop accuracy is provided by the {\tt LHAPDF6} library \cite{Buckley:2014ana} including five active flavors. Our default renormalization and factorization scales are chosen as $\mu=M_{HH}/2$, where $M_{HH}$ is the invariant mass of the produced Higgs pair. 

The phase space integration in Eq.~\eqref{eq:Xsection} is carried out and optimized using {\tt{Parni}} \cite{vanHameren:2007pt}. A total of $3\times10^5$ events are generated at the LO, and the results are cross-checked with {\tt{MadGraph5}} \cite{Alwall:2014hca}. The LO events are stored in the form of modified Les Houches event files \cite{Alwall:2006yp}, enabling reweighting to the NLO events.

We calculate $F_1$ and $F_2$ at NLO for $1.8\times 10^4$ reweighted events. This enables us to compute, on one hand, the NLO cross sections and, on the other hand, the ${\cal K}$ factor. The ${\cal K}$ factor is defined as the ratio of (differential) cross sections at NLO to that at LO, representing the main focus of this Letter. To compute the ${\cal K}$ factor, we utilize these $1.8\times 10^4$ reweighted NLO events and the corresponding LO events.  We have verified that  ${\cal K}$ factors computed using this method align with those computed from uncorrelated events.

\begin{table}[htbp]
\begin{center}
\renewcommand\arraystretch{1.8}
\begin{tabular}{ccccccc}
\hline
\hline
$\mu$  & $M_{HH}/2$  & $\sqrt{p_{T}^2+m_H^2}$ & $m_H$ \\
\hline
LO   & 19.96(6)   & 21.11(7)  & 25.09(8) \\
NLO  & 19.12(6)   & 20.21(6)  & 23.94(8) \\
${\cal K}$ factor  & 0.958(1)   &     0.957(1)   &   0.954(1)     \\
\hline
\hline
\end{tabular}
\caption{LO and NLO ${\rm EW}$ corrected integrated cross sections (in fb) with $\sqrt{s}= 14$ TeV based on $1.8\times 10^4$ reweighted events. The uncertainties arise from statistical errors in phase space integration.}
\label{total-cs}
\end{center}
\end{table}

It is important to emphasize that, although there are alternative choices for the aforementioned parameters such as PDF or renormalization scale, they have negligible impact on the ${\cal K}$ factor. To illustrate this point, Table \ref{total-cs} presents results for $\mu=m_H$ and $\mu=\sqrt{p_{T}^2+m_H^2}$ in addition to $\mu=M_{HH}/2$. The observed differences with varying scale choices are around $20\%$ both at LO and NLO, attributed to the $\mu$ dependence of the strong coupling $\alpha_s$ and gluon PDF $f_{g/p}$. This dependence can be mitigated after incorporating higher-order QCD corrections \cite{Jones:2023uzh}. In contrast, the ${\cal K}$ factor remains rather stable for different choices of $\mu$, as expected.

\sect{Differential cross sections}
Differential cross sections offer a wealth of information about physics, both within the SM and in scenarios beyond it. The impact of NLO EW corrections can vary across specific regions of the phase space compared to the full phase space.

As indicated in Table~\ref{total-cs}, the statistical uncertainty for the ${\cal K}$ factor is smaller than that of the NLO cross section. This discrepancy arises because the differential ${\cal K}$ factor exhibits a much flatter behavior compared to the differential cross section, enabling the former to get a controllable error with far fewer events for numerical integration.
Given that the computation at LO is significantly more economical, we proceed to compute NLO differential cross sections using the following relation:
\begin{align}
    \Delta\sigma^{\text{NLO}} = \Delta {\cal K} \Delta\sigma^{\text{LO}},
\end{align}
where $\Delta {\cal K}$ is the ${\cal K}$ factor calculated in a specific phase space region using the same events at LO and NLO, and $\Delta\sigma^{\text{LO}}$ is the LO result computed in the same region but using a significantly larger number of events.

With the $1.8\times 10^4$ reweighted events, we can compute the ${\cal K}$ factor quite accurately for most bins, except for those with very large $M_{HH}$ or $p_{T}$. For each of these bins, we compute an additional 400 reweighted events and use them to determine the corresponding ${\cal K}$ factor.

\begin{figure}[htbp]
    \includegraphics[width=0.4\textwidth, trim = 1 0cm 1 5cm, clip]{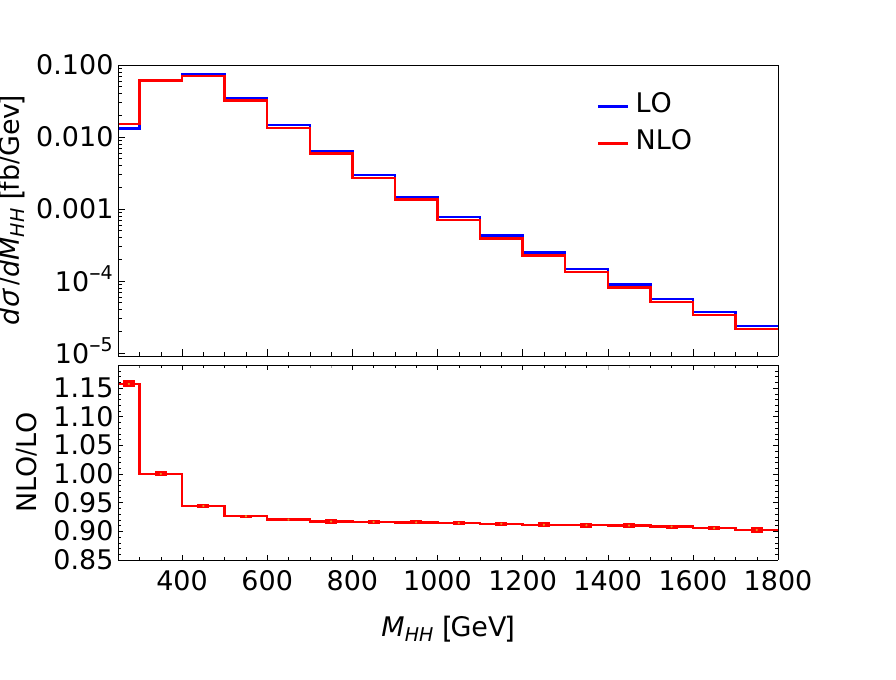}
\caption{\label{mhh}
Invariant mass distribution of the Higgs pair  with $\sqrt{s}= 14$ TeV. The upper plot shows absolute predictions, and the lower panel displays the differential ${\cal K}$ factor with error bars representing statistical errors.}
\end{figure}

\begin{figure}[htbp]
\begin{center}
    \includegraphics[width=0.4\textwidth, trim = 1 0cm 1 5cm, clip]{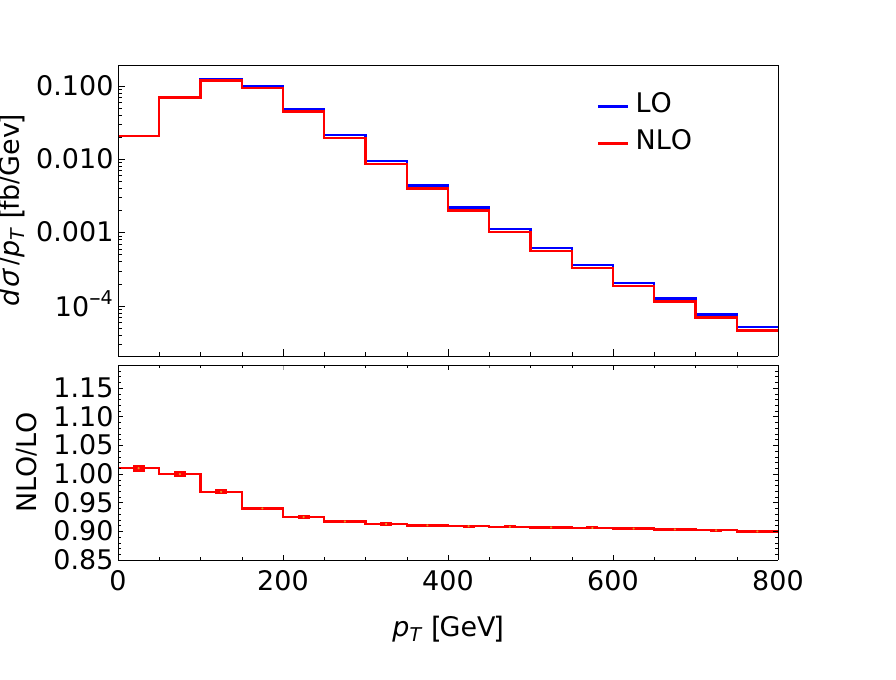}
\caption{\label{pth}
%Transverse momentum distribution of one of the two Higgs bosons  with $\sqrt{s}= 14$ TeV. The upper plot shows absolute predictions, and the lower panel displays the differential ${\cal K}$-factor with error bars representing statistical errors.
Same as Fig. \ref{mhh}, but for transverse momentum distribution of one of the two Higgs bosons.}
\end{center}
\end{figure}

In Fig. \ref{mhh}, we present the $M_{HH}$ distribution. A significant positive correction of approximately $+15\%$ is observed in the first bin. In fact, we find that the EW correction for phase space points near the $HH$ production threshold can exceed $+70\%$.  A similar result has also been obtained in Ref. \cite{Muhlleitner:2022ijf} where top-Yukawa corrections have been considered, partially in the heavy top-quark mass limit. 
This can be understood by examining the EW corrections using heavy top-quark mass expansion. As shown in Ref. \cite{Davies:2023npk}, the leading term in the expansion at NLO is larger than that at LO by $m_t^4$, which explains the substantial increase near threshold. However, above the threshold, the expansion becomes unreliable, and consequently, the enhancement should no longer exist. Indeed, as $M_{HH}$ increases, the ${\cal K}$ factor decreases dramatically initially and then slows down as it moves away from the threshold. 
%Besides, the EW correction at $M_{HH}$ tail reported in Ref. \cite{Muhlleitner:2022ijf} tend to be 0 in contrary to -10\% in our work. 
The pattern is similar for the $p_{T}$ distribution in Fig. \ref{pth}, where the correction is positive initially and subsequently becomes negative. In regions of either large $M_{HH}$ or large $p_{T}$, we find the NLO EW correction to be approximately $-10\%$. We explicitly checked phase space points with $\sqrt{\hat{s}}$ close to $14$ TeV and found the corrections to be as substantial as $-30\%$ at the matrix element squared level. However, the gluon luminosity is highly suppressed in this region, and thus, it does not contribute significantly to (differential) cross sections.

\begin{figure}[htbp]
\begin{center}
    \includegraphics[width=0.4\textwidth, trim = 1 0cm 1 5cm, clip]{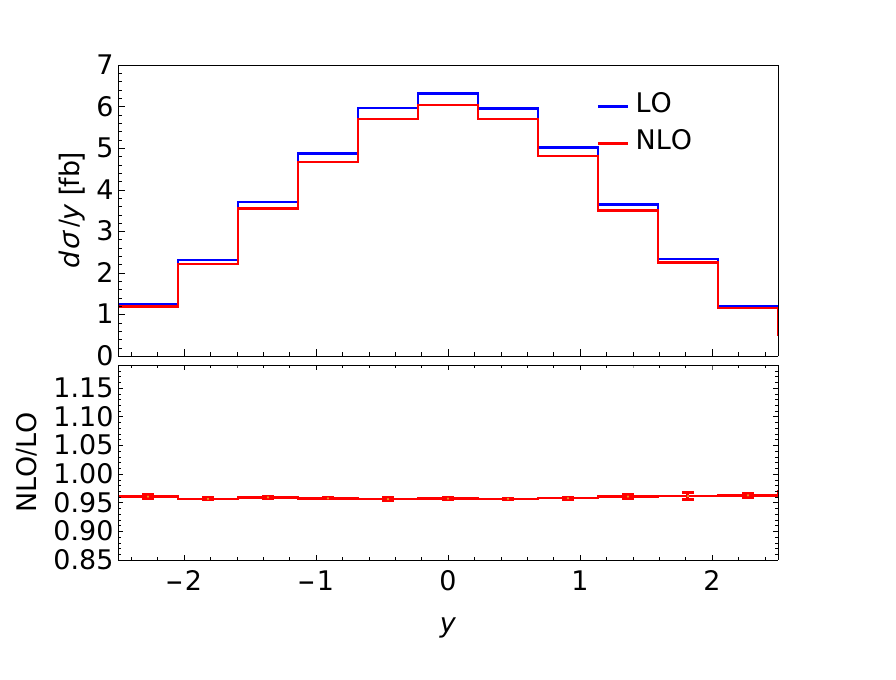}
\caption{\label{yh} 
%Rapidity distribution of one of the two Higgs bosons  with $\sqrt{s}= 14$ TeV. The upper plot shows absolute predictions, and the lower panel displays the differential ${\cal K}$-factor with error bars representing statistical errors.
Same as Fig. \ref{mhh}, but for rapidity distribution distribution of one of the two Higgs bosons.}
\end{center}
\end{figure}

In Fig. \ref{yh}, we display the rapidity distribution of one of the two Higgs bosons. A nearly flat ${\cal K}$ factor is observed, approximately 0.96, similar to the total cross section.

\sect{Summary}
Double Higgs production is considered the golden channel to probe the Higgs self-coupling, with NLO EW corrections representing the most substantial source of theoretical uncertainties. In this Letter, we, for the first time, compute the complete NLO EW corrections for the production of double Higgs bosons in the dominant gluon-gluon fusion channel at the LHC. The NLO EW corrections are approximately $-4\%$ at the total cross-section level, proving to be insensitive to the choice of parameters. For the differential cross sections, the EW corrections can be significant in certain phase space regions. Specifically, for dimensionful observables, EW corrections reach up to $+15\%$ at the beginning of the spectrum and $-10\%$ in the tail. For dimensionless observable, the rapidity distribution, the NLO EW corrections are flat and approximately $-4\%$. The complete NLO EW corrections are found to be close to the size of $\rm N^3LO$ QCD corrections  in heavy top-quark mass limit \cite{Chen:2019lzz}, which amounts to 8\%.

As emphasized in the Introduction, the size of NLO EW corrections constituted the primary source of theoretical uncertainties in this process before this Letter. Our Letter has substantially reduced the theoretical uncertainties by addressing this critical aspect. 
The remaining theoretical uncertainties at present primarily stem from high-order QCD effects and QCD parameters, accounting for approximately 3\% \cite{Jones:2023uzh}.  Upon future comparisons of such precise predictions with experimental measurements, the standard model can be further confirmed if agreement is observed, or potential beyond-the-standard-model effects may be indicated if any tension arises.

%%%%%%%%%%%%%%%%%%%%%%%%%%%%%%%%%%%%%%%%%%%%%%%%%
\begin{acknowledgments}
We thank X. Chen, X. Guan and J. Wang for many helpful discussions. We also would like to thank H. Zhang and X. Zhang for helpful discussions in comparing our findings with theirs. The work was supported in part by the National Natural Science Foundation of China (Grants No. 12325503 and No. 11975029), the National Key Research and Development Program of China under	Contract No.~2020YFA0406400, and the China Postdoctoral Science
Foundation under Grants No. 2022TQ0012 and No. 2023M730097.
We acknowledge the computational support from the High-performance Computing Platform of Peking University. Feynman diagrams are drawn using the {\tt FeynGame} program \cite{Harlander:2020cyh}.
\end{acknowledgments}

%%%%%%%%%%%%%%%%%%%%%%%%%%%%%%%%%%%%%%%%%%%%%%%%%

\providecommand{\href}[2]{#2}\begingroup\raggedright\endgroup

%\bibliographystyle{utphysMa}
%\bibliography{ref}

\end{document}